\documentclass[preprint,showpacs,preprintnumbers,amsmath,amssymb,pra,superscriptaddress]{revtex4}

\usepackage[dvips]{graphicx} 
\usepackage{amsfonts}
\usepackage{amscd}
\usepackage{amsmath}    

\begin{document}

\title{Quantum key distribution with entangled photon sources}

\author{Xiongfeng Ma}
\email{xima@physics.utoronto.ca}
\author{Chi-Hang Fred Fung}
\email{cffung@comm.utoronto.ca}
\author{Hoi-Kwong Lo}
\email{hklo@comm.utoronto.ca}
\affiliation{%
Center for Quantum Information and Quantum Control,\\
Department of Electrical \& Computer Engineering and Department of Physics, \\
University of Toronto, Toronto, Ontario, Canada, M5S 1A7 \\
}%

\begin{abstract}
A parametric down-conversion (PDC) source can be used as either a triggered single
photon source or an entangled photon source in quantum key distribution (QKD). The
triggering PDC QKD has already been studied in the literature. On the other hand, a
model and a post-processing protocol for the entanglement PDC QKD are still missing.
In this paper, we fill in this important gap by proposing such a model and a
post-processing protocol for the entanglement PDC QKD. Although the PDC model is
proposed to study the entanglement-based QKD, we emphasize that our generic model
may also be useful for other non-QKD experiments involving a PDC source. Since an
entangled PDC source is a basis independent source, we apply Koashi-Preskill's
security analysis to the entanglement PDC QKD. We also investigate the entanglement
PDC QKD with two-way classical communications. We find that the recurrence scheme
increases the key rate and Gottesman-Lo protocol helps tolerate higher channel
losses. By simulating a recent 144km open-air PDC experiment, we compare three
implementations --- entanglement PDC QKD, triggering PDC QKD and coherent state QKD.
The simulation result suggests that the entanglement PDC QKD can tolerate higher
channel losses than the coherent state QKD. The coherent state QKD with decoy states
is able to achieve highest key rate in the low and medium-loss regions. By applying
Gottesman-Lo two-way post-processing protocol, the entanglement PDC QKD can tolerate
up to 70dB combined channel losses (35dB for each channel) provided that the PDC
source is placed in between Alice and Bob. After considering statistical
fluctuations, the PDC setup can tolerate up to 53dB channel losses.
\end{abstract}

\maketitle

\section{Introduction}
There are mainly two types of quantum key distribution (QKD) schemes. One is the
prepare-and-measure scheme such as BB84 \cite{BB_84} and the other is the
entanglement based QKD such as Ekert91 \cite{Ekert_91}. The security of both types
of QKD has been proven in the last decade, see for example,
\cite{Mayers_01,LoChauQKD_99,ShorPreskill_00}. For a review of quantum cryptography,
one may refer to \cite{GRTZ_02}. Meanwhile, researchers have also proven the
security of QKD with realistic devices, such as
\cite{MayersYao_98,IndividualAttack_00,BLMS_00,FGSZ_01,ILM_01,KoashiPreskill_03,GLLP_04}.

In the original proposal of the BB84 protocol, a single photon source is required. Unfortunately,
single photon sources are still not commercially available. Instead, a weak coherent state source
is widely used as an imperfect single photon source. Throughout this paper, we call this
implementation as coherent state QKD. Many coherent state QKD experiments have been performed since
the first QKD experiment \cite{BBBSS_92}, see for example
\cite{Townsend_98,RGGGZ_98,BGKHJTLS_99,KZHWGTR_Free23_02,GYS_04,MZHGG_Beijing_05}.

Decoy state method \cite{Hwang_03} has been proposed as a useful method for substantially improving
the performance of the coherent state QKD. The security of QKD with decoy states has been proven
\cite{LoDecoy_03,MasterReport,Decoy_05}. Asymptotically, the coherent state QKD with decoy states
is able to operate as good as QKD with perfect single photon sources in the sense that the key
generation rates given by both setups linearly depend on the channel transmittance \cite{Decoy_05}.
Afterwards, some practical decoy-state protocols are proposed
\cite{Wang_05,HEHN_05,Wang2_05,Practical_05}. The experimental demonstrations for decoy state
method have been done recently
\cite{ZQMKQ_06,ZQMKQ60km_06,LosAlamosDecoy_07,PDC144_07,PanDecoy_07,YSS_Decoy07}. Other than decoy
state method, there are other approaches to enhance the performance of the coherent state QKD, such
as QKD with strong reference pulses \cite{Koashi_04,TLMB_Strong06} and differential-phase-shift QKD
\cite{IWY_DPS02}.

Besides the coherent source, there is another source that can be used for QKD --- parametric
down-conversion (PDC) source. With a PDC source, one can realize either prepare-and-measure or
entanglement-based QKD protocols. To implement a prepare-and-measure QKD protocol, one can use a
PDC source as a triggered single photon source. To implement an entanglement-based QKD protocol, on
the other hand, one can use the polarization entanglement between two modes of the light emitted
from a PDC source. We call these two implementations triggering PDC QKD and entanglement PDC QKD.
With a entangled source, one can also implement QKD protocols based on causality
\cite{MW_Causality_06} and Bell's inequality \cite{AGM_Bell_06}.

The model and post-processing of the triggering PDC QKD have already been studied
\cite{IndividualAttack_00}. Recently, there are some practical decoy state proposals for the
triggering PDC QKD \cite{MauererSilberhorn_06,AYKI_06,WWG_07}. In this paper, we will focus on the
asymptotic decoy state protocol \cite{Decoy_05}, which is the upper bound of all these practical
decoy state protocols when threshold detectors are used by Alice and Bob.

On the other hand, the model and post-processing for the entanglement PDC QKD are still missing. In
this paper, we present a model for the entanglement PDC QKD. From the model, we find that an
entangled PDC source is a basis independent source for QKD. Based on this observation, we propose a
post-processing scheme by applying Koashi-Preskill's security analysis \cite{KoashiPreskill_03}.

Recently, a free-space distribution of entangled photons over 144 km has been demonstrated
\cite{PDC144_06}. We simulate this experiment setup and compare three QKD implementations:
entanglement PDC QKD, triggering PDC QKD and coherent state QKD. In the simulation, we also apply
Gottesman-Lo two-way post-processing protocol \cite{TwoWay_03} and a recurrence scheme
\cite{GVV_05}, see also \cite{TwoWay_06}.

The main contributions of this paper are:
\begin{itemize}
\item
We present a model for the entanglement PDC QKD. Although the model is proposed to
study the entanglement-based QKD, this generic model may also be useful for other
non-QKD experiments involving a PDC source.

\item
From the model, we find that an entangled PDC source is a basis independent source
for QKD. Based on this observation, we propose a post-processing scheme for the
entanglement PDC QKD. Essentially, we apply Koashi-Preskill's security analysis
\cite{KoashiPreskill_03}.

\item
By simulating a PDC experiment \cite{PDC144_06}, we compare three QKD
implementations: entanglement PDC QKD, triggering PDC QKD and coherent state QKD. In
the entanglement PDC QKD, we consider two cases: source in the middle and source at
Alice's side.

\item
In the case that the PDC source is placed in between Alice and Bob, we find that the
entanglement PDC QKD can tolerate the highest channel losses, up to 70 dB by
applying Gottesman-Lo two-way classical communication post-processing protocol
\cite{TwoWay_03}. We remark that a 35dB channel loss is comparable to the estimated
loss in a satellite to ground transmission in the literature
\cite{Zeilinger_Satellite_03, Zeilinger_SpaceGround_04}.


\item
We consider statistical fluctuations for the entanglement PDC QKD. In this case, the
PDC setup can tolerate up to 53dB channel loss.


\item
The coherent state QKD with decoy states is able to achieve highest key rate in the
low- and medium-loss regions.
\end{itemize}

In Section \ref{Impl}, we will review two experiment setups of the entanglement PDC
QKD. In Section \ref{Model}, we will model the entanglement PDC QKD. In Appendix
\ref{QBER}, we calculate the quantum bit error rate in the entanglement PDC QKD. In
Section \ref{Post}, we will propose a post-processing scheme for the entanglement
PDC QKD. In Section \ref{Simulation}, we will compare the entanglement PDC QKD, the
triggering PDC QKD and the coherent state QKD by simulating a real PDC experiment.
We also apply protocols based on two-way classical communications and consider
statistical fluctuations. In Appendix \ref{Optimalmu}, we investigate the optimal
$\mu$ for the entanglement PDC QKD.

\section{Implementation} \label{Impl}
In general, the PDC source does not necessarily belong to one of the two legitimate users of QKD,
Alice or Bob. One can even assume that an eavesdropper, Eve, owns the PDC source. In this section
we will compare two experimental setups of the entanglement PDC QKD due to the position of the PDC
source, in between Alice and Bob or at Alice's side.

Let us start with a general discussion about an entangled PDC source. With the
rotating-wave approximation, the Hamiltonian of the PDC process can be written as
\cite{KokBraunstein_00}
\begin{equation}\label{Impl:HamiltonianEn}
\begin{aligned}
H &= i\chi(a^\dagger_Hb^\dagger_V-a^\dagger_Vb^\dagger_H)+h.c. \\
\end{aligned}
\end{equation}
where $h.c.$ means Hermitian conjugate and $\chi$ is a coupling constant depending
on the crystal nonlinearity and the amplitude of the pump beam. The operators $a_i$
and $b_i$ are the annihilation operators for rectilinear polarizations $i\in\{H,V\}$
in mode $a$ and $b$ respectively. Mode $a$ and mode $b$ are the modes sent to Alice
and Bob, respectively.


In the Section \ref{Model}, we will focus on modeling the measurement of the rectilinear
polarization ($Z$) basis. Due to symmetry, all the calculations can be applied to $X$ basis too.

\subsection{Source in the middle}
\begin{figure}[hbt]
\centering \resizebox{12cm}{!}{\includegraphics{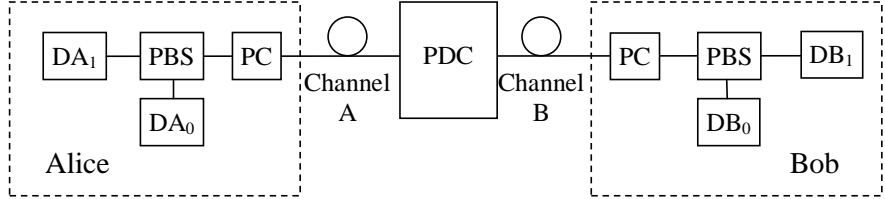}} \caption{A schematic
diagram for the entanglement PDC QKD. Alice and Bob connect to a entangled PDC
source by optical links. They each receives one of two entangled modes coming out
from the PDC source. Both Alice and Bob randomly choose basis (by polarization
controllers) to measure the arrived signals (by single photon detectors). PC:
polarization controller; PBS: polarized beam splitter; DA$_0$, DA$_1$, DB$_0$,
DB$_1$: threshold detectors.} \label{Fig:PDCen}
\end{figure}

First we consider the case that the PDC source sits in between Alice and Bob. The schematic diagram
is shown in FIG.~\ref{Fig:PDCen}.

As shown in FIG.~\ref{Fig:PDCen}, a PDC source provides two entangled modes, $a$ and $b$, which are
sent to Alice and Bob, respectively. After receiving the signals, Alice and Bob each randomly
chooses a basis ($X$ or $Z$) to perform a measurement. One key observation of this setup is that
the state emitted from the PDC source is independent of the bases Alice and Bob choose for the
measurements.

\subsection{Source at Alice's side} \label{AliceSide}
\begin{figure}[hbt]
\centering \resizebox{12cm}{!}{\includegraphics{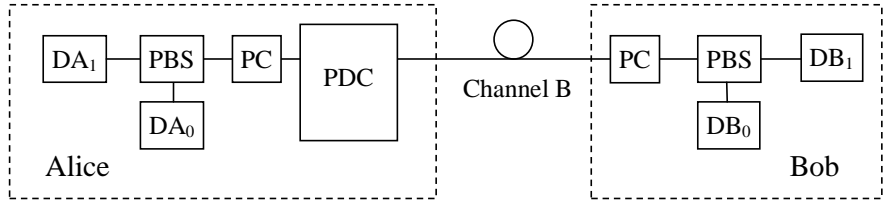}} \caption{A schematic diagram for the
entanglement PDC QKD. Alice measures one of entangled modes coming out from the PDC source and
sends Bob the other mode.} \label{Fig:PDCenA}
\end{figure}

Another case is that Alice owns the PDC source. The schematic diagram is shown in
FIG.~\ref{Fig:PDCenA}.

Compare FIG.~\ref{Fig:PDCen} and \ref{Fig:PDCenA}, we can see that the only difference is the
position of the PDC source. As we will see in Section \ref{Post}, the post-processing these two
setups are similar.

We remark that in the second setup, Alice's measurement commutes with Bob's measurement.
Thus, we have the same observation as before that the PDC source state is independent of the
measurement bases.

Therefore, for both setups the entangled PDC source is a basis-independent source. It follows that
the entanglement PDC QKD is a basis independent QKD.

\section{Model} \label{Model}
In this section, we will model entangled PDC sources, channel and detectors for the
entanglement PDC QKD. We emphasize that this model is applicable for both experiment
setups described in Section \ref{Impl}.

\subsection{An entangled PDC source}
From Eq.~\eqref{Impl:HamiltonianEn}, the state emitted from a type-II PDC source can
be written as \cite{KokBraunstein_00}
\begin{equation}\label{Model:PDCstate}
\begin{aligned}
|\Psi\rangle=(\cosh\chi)^{-2}\sum_{n=0}^{\infty}\sqrt{n+1}\tanh^n\chi|\Phi_n\rangle,
\end{aligned}
\end{equation}
where $|\Phi_n\rangle$ is the state of an $n$-photon-pair, given by
\begin{equation}\label{Model:PDCn}
\begin{aligned}
|\Phi_n\rangle=\frac{1}{\sqrt{n+1}}\sum_{m=0}^{n}(-1)^m|n-m,m\rangle_a|m,n-m\rangle_b.
\end{aligned}
\end{equation}
For example, when $n=1$, Eq.~\eqref{Model:PDCn} will give a Bell state,
\begin{equation}\label{Model:EPR}
\begin{aligned}
|\Phi_1\rangle &= \frac{1}{\sqrt{2}}(|1,0\rangle_a|0,1\rangle_b-|0,1\rangle_a|1,0\rangle_b) \\
               &= \frac{1}{\sqrt{2}}(|\leftrightarrow\rangle_a|\updownarrow\rangle_b-|\updownarrow\rangle_a|\leftrightarrow\rangle_b), \\
\end{aligned}
\end{equation}
Here we use the polarizations $|1,0\rangle=|\leftrightarrow\rangle$ and
$|0,1\rangle=|\updownarrow\rangle$ as a qubit basis (Z basis) for QKD. From
Eq.~\eqref{Model:PDCstate}, the probability to get an $n$-photon-pair is
\begin{equation}\label{Model:Pn}
\begin{aligned}
P(n)=\frac{(n+1)\lambda^n}{(1+\lambda)^{n+2}}
\end{aligned}
\end{equation}
where we define $\lambda\equiv\sinh^2\chi$. The expected photon pair number is $\mu=2\lambda$,
which is the average number of photon pairs generated by one pump pulse, characterizing the
brightness of a PDC source.


\subsection{Detection} \label{Detection}
We assume that the detection probabilities of the photons in the state of Eq.~\eqref{Model:PDCn}
are independent. Define $\eta_A$ and $\eta_B$ to be the detection efficiencies for Alice and Bob,
respectively. Both $\eta_A$ and $\eta_B$ take into account of the channel losses, detector
efficiencies, coupling efficiencies and losses inside the detector box. For an $n$-photon-pair, the
overall transmittance is
\begin{equation}\label{Model:etan}
\begin{aligned}
\eta_n=[1-(1-\eta_A)^n][1-(1-\eta_B)^n].
\end{aligned}
\end{equation}
We remark that the channel loss is included in $\eta_A$ and $\eta_B$. Thus, this
model can be applied to either of following two cases: 1) the PDC source is in
between Alice and Bob or 2) the PDC source is at Alice (or Bob)'s side.

\textbf{Yield:} define $Y_n$ to be the yield of an $n$-photon-pair, i.e., the conditional probability of a coincidence
detection event given that the PDC source emits an $n$-photon-pair. 
$Y_n$ mainly comes from two parts, the background and the true signal. Assuming that the background
counts are independent of the signal photon detection, then $Y_n$ is given by
\begin{equation}\label{Model:Yn}
\begin{aligned}
Y_n 
    &= [1-(1-Y_{0A})(1-\eta_A)^n][1-(1-Y_{0B})(1-\eta_B)^n] \\
\end{aligned}
\end{equation}
where $Y_{0A}$ and $Y_{0B}$ are the background count rates at Alice's and Bob's sides,
respectively. The vacuum state contribution is $Y_0=Y_{0A}Y_{0B}$.
The {\it gain} of the $n$-photon-pair $Q_n$, which is the product of Eqs.~\eqref{Model:Pn} and
\eqref{Model:Yn}, is given by
\begin{equation}\label{Model:Qn}
\begin{aligned}
Q_n &= Y_nP(n) \\
    &= [1-(1-Y_{0A})(1-\eta_A)^n][1-(1-Y_{0B})(1-\eta_B)^n]\frac{(n+1)\lambda^n}{(1+\lambda)^{n+2}}. \\
\end{aligned}
\end{equation}

The overall gain is given by
\begin{equation}\label{Model:Gain}
\begin{aligned}
Q_{\lambda} &= \sum_{n=0}^{\infty} Q_n \\
        &= 1-\frac{1-Y_{0A}}{(1+\eta_A\lambda)^2}-\frac{1-Y_{0B}}{(1+\eta_B\lambda)^2}+\frac{(1-Y_{0A})(1-Y_{0B})}{(1+\eta_A\lambda+\eta_B\lambda-\eta_A\eta_B\lambda)^2}.
\end{aligned}
\end{equation}
Here the overall gain $Q_\lambda$ is the probability of a coincident detection event given a pump
pulse. Note that the parameter $\lambda$ is one half of the expected photon pair number $\mu$.

The overall quantum bit error rate (QBER, $E_\lambda$) is given by
\begin{equation}\label{Model:QBER}
\begin{aligned}
E_{\lambda}Q_{\lambda} 
               =&e_0Q_{\lambda}-\frac{2(e_0-e_{d})\eta_A\eta_B\lambda(1+\lambda)}{(1+\eta_A\lambda)(1+\eta_B\lambda)(1+\eta_A\lambda+\eta_B\lambda-\eta_A\eta_B\lambda)} \\
\end{aligned}
\end{equation}
where $Q_{\lambda}$ is the gain given in Eq.~\eqref{Model:Gain}. The calculation of the
$E_{\lambda}$ is shown in Appendix \ref{QBER}.

\section{Post-processing} \label{Post}
As mentioned in Section \ref{Impl}, the entanglement PDC QKD is a basis-independent
QKD. Thus, we can apply Koashi and Preskill's security proof
\cite{KoashiPreskill_03}. The key generation rate is given by
\begin{equation} \label{Post:KeyEn}
R \geq q \{Q_{\lambda}[1-f(\delta_b)H_2(\delta_b)-H_2(\delta_p)]\}.
\end{equation}
where $q$ is the basis reconciliation factor (1/2 for the BB84 protocol due to the fact that half
of the time Alice and Bob disagree with the bases, and if one uses the efficient BB84 protocol
\cite{EffBB84_05}, $q\approx1$), the subscript $\lambda$ denotes for one half of the expected
photon number $\mu$, $Q_{\lambda}$ is the overall gain, $\delta_b$ ($\delta_p$) is the bit (phase)
error rate, $f(x)$ is the bi-direction error correction efficiency (see, for example,
\cite{BrassardSalvail_93}) as a function of error rate, normally $f(x)\ge1$ with Shannon limit
$f(x)=1$, and $H_2(x)$ is the binary entropy function,
$$
H_2(x)=-x\log_2(x)-(1-x)\log_2(1-x).
$$

Due to the symmetry of $X$ and $Z$ bases measurements, as shown in Section
\ref{Impl}, $\delta_b$ and $\delta_p$ are given by
\begin{equation} \label{Post:deltabp}
\delta_b=\delta_p=E_\lambda,
\end{equation}
where $E_\lambda$ is the overall QBER. This equation is true for the asymptotic
limit of an infinitely long key distribution. Later in the Subsection of
\ref{StaFlu}, we will see Eq.~\eqref{Post:deltabp} may not be true when statistical
fluctuations are taken into account.

We remark that in Koashi and Preskill's security proof, the squash model
\cite{GLLP_04} is applied. In the squash model, Alice and Bob project the state onto
the qubit Hilbert space before $X$ or $Z$ measurements. For more details of the
squash model, one can refer to \cite{GLLP_04}. In the case that Alice owns the PDC
source, as discussed in Subsection \ref{AliceSide}, the key rate formula of
Eq.~\eqref{Post:KeyEn} has been proven \cite{Koashi_NewModel_06} to be valid for the
QKD with threshold detectors without the squash model, see also
\cite{LoPreskill_NonRan_06}.  We also notice that this post-processing scheme,
Eqs.~\eqref{Post:KeyEn} and \eqref{Post:deltabp}, can also be derived from the
security analysis based on the uncertainty principle \cite{Koashi_06}.

In Eq.~\eqref{Post:KeyEn}, $Q_\lambda$ can be directly measured from a QKD experiment and
$E_\lambda$ can be estimated by error testing. In the simulation shown in Section \ref{Simulation},
we will use Eqs.~\eqref{Model:Gain} and \eqref{Model:QBER}.

We remark that the post-processing for the entanglement PDC QKD is simpler than the
coherent state QKD and triggering PDC QKD. In the entanglement PDC QKD, all the
parameters needed for the post-processing ($Q_{\lambda}$ and $E_{\lambda}$) can be
directly calculated or tested from the measured classical data. In the coherent PDC
QKD and the triggering PDC QKD, on the other hand, Alice and Bob need to know the
value of some experimental parameters ahead, such as the expected photon number
$\mu$ ($=2\lambda$), and also need to estimate the gain and error rate of the single
photon states $Q_1$ and $e_1$, which make the statistical fluctuation analysis
difficult \cite{Practical_05}.

The post-processing can be further improved by introducing two-way classical
communications between Alice and Bob \cite{TwoWay_03,TwoWay_06}. Also, the adding
noise technique may enhance the performance \cite{KGR_noise_05}.

\section{Simulation} \label{Simulation}
In this section, we will first compare three QKD implementations: entanglement PDC
QKD, triggering PDC QKD and coherent state QKD. Then we will apply post-processing
protocols with two-way classical communications to the entanglement PDC QKD.
Finally, we will consider statistical fluctuations.

We deduce experimental parameters from Ref.~\cite{PDC144_06} due to the model given
in Subsection \ref{Model}, which are listed in TABLE \ref{Tab:PDC144}. For the
coherent state QKD, we use $\eta_A=1$ since Alice prepares the states in this case.
In the following simulations, we will use $q=1/2$ and $f(E_\mu)=1.22$
\cite{BrassardSalvail_93}.

\begin{table}[hbt]
\centering
\begin{tabular}{|c|c|c|c|c|c|c|c|c|c|c|} \hline
Repetition rate & Wavelength & $\eta_{Alice}$ & $\eta_{Bob}$ & $e_{d}$ & $Y_0$ \\
\hline
249MHz & 710 nm & 14.5\% & 14.5\% & 1.5\% & $6.02\times 10^{-6}$ \\
\hline
\end{tabular}
\caption{Experimental parameters deduced from 144 km PDC experiment \cite{PDC144_06}. Here we
assume Alice and Bob use detectors with same characteristics. $e_d$ is the intrinsic detector error
rate. $Y_0$ is the background count rate. $\eta_{Alice}$ ($\eta_{Bob}$) is the detection efficiency
in Alice (Bob)'s box, including detector efficiency and internal optical losses. The overall
transmittance $\eta_A$ ($\eta_B$) is the products of Alice (Bob)'s channel transmission efficiency
and $\eta_{Alice}$ ($\eta_{Bob}$).} \label{Tab:PDC144}
\end{table}

The optimal expected photon number $\mu$ of the coherent state QKD has been
discussed in Ref.~\cite{IndividualAttack_00,Practical_05}. In Appendix
\ref{Optimalmu}, we investigate the optimal $\mu$ ($2\lambda$) for the entanglement
PDC QKD. Not surprisingly, we find that the optimal $\mu$ for the entanglement PDC
QKD is in the order of 1, $\mu=2\lambda=O(1)$. Thus, the key generation rate given
in Eq.~\eqref{Post:KeyEn} depends linearly on the channel transmittance.

\subsection{Comparison of three QKD implementations}
\begin{figure}[hbt]
\centering \resizebox{12cm}{!}{\includegraphics{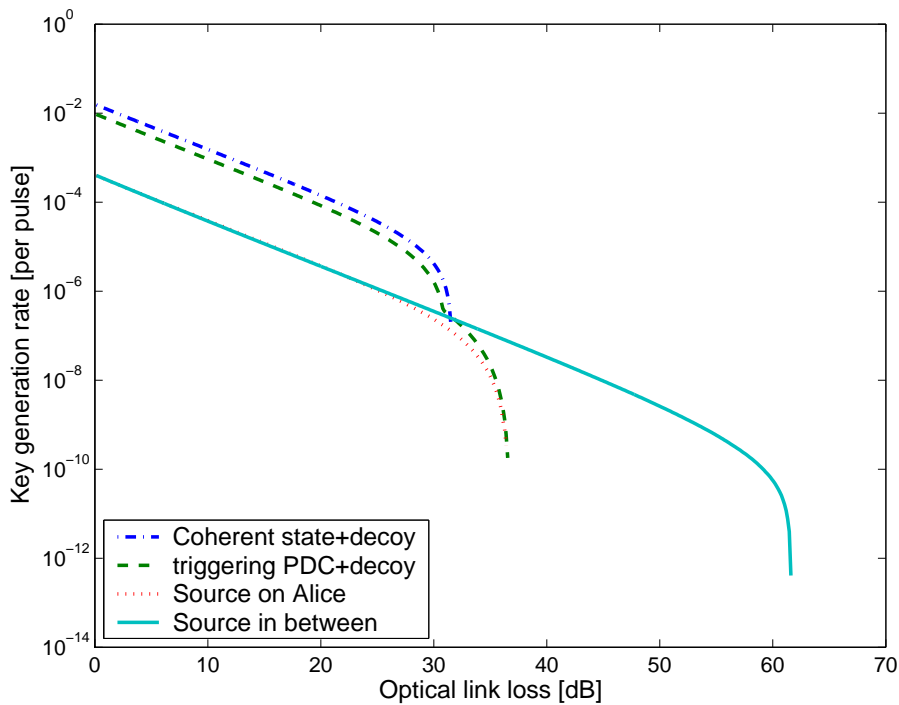}} \caption{(Color online)
Plot of the key generation rate in terms of the optical loss, comparing comparing
four cases: coherent state QKD+aysmptotic decoy, triggering PDC+asymptotic decoy,
and entanglement PDC QKD (source in the middle and source at Alice's side). For the
coherent state QKD+decoy, we use $\eta_A=1$. We numerically optimize $\mu$
($2\lambda$) for each curve.} \label{Fig:Toy}
\end{figure}

In the first simulation, we assume that Alice is able to adjust the expected photon
pair number $\mu$ ($2\lambda$, the brightness of the PDC source) in the region of
$[0,1]$. Thus, we can optimize $\mu$ for the entanglement PDC QKD and the triggering
PDC QKD. The simulation results are shown in FIG.~\ref{Fig:Toy}.

From FIG.~\ref{Fig:Toy}, we have the following remarks.
\begin{enumerate}
\item
The entanglement PDC QKD can tolerate the highest channel losses in the case that
the source is placed in middle between Alice and Bob.

\item
The coherent state QKD with decoy states is able to achieve the highest key rate in
the low- and medium-loss region. This is because in the coherent state QKD
implementation, Alice does not need to detect any photons, which will effectively
give $\eta_A=1$ in the PDC QKD implementations.

\item
Comparing two cases of the entanglement PDC QKD: source in the middle and source at
Alice's side, they yields similar key rate in the low- and media- region. But source
in the middle case can tolerate higher channel losses.
\end{enumerate}

In the following simulations, we will focus on the case that the entangled PDC
source sits in the middle between Alice and Bob.

\subsection{With two-way classical communications}
\begin{figure}[hbt]
\centering \resizebox{12cm}{!}{\includegraphics{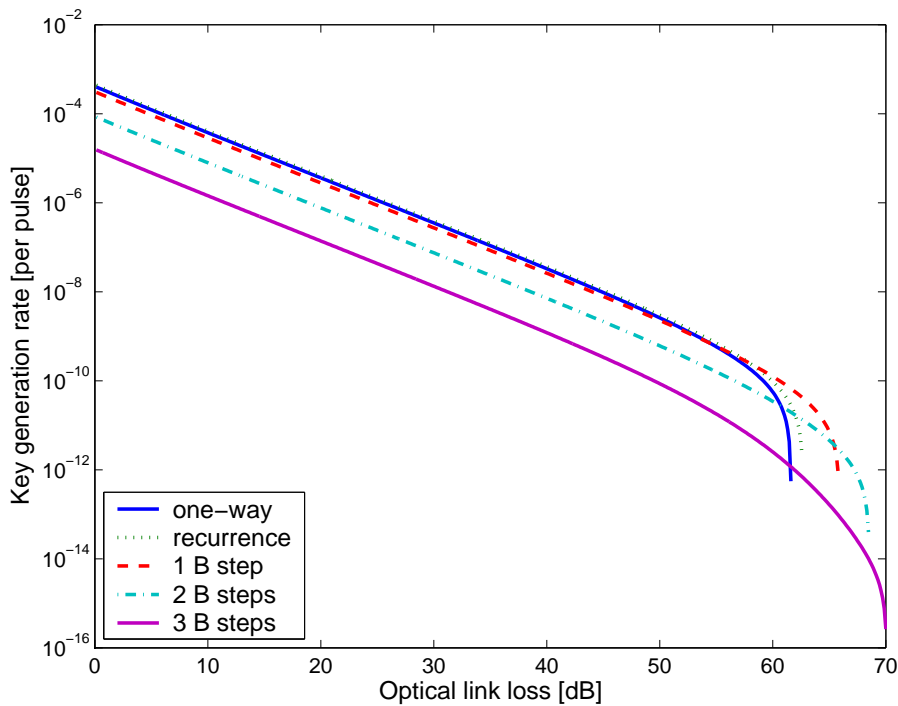}} \caption{(Color online)
Plot of the key generation rate in terms of the optical loss. We apply the
recurrence idea and up to 3 B steps. $\mu$ is numerically optimized for each curve.}
\label{Fig:TwoWay}
\end{figure}

We can also apply the idea of post-processing with two-way classical communications.
Similar to the argument of Ref.~\cite{TwoWay_06}, we combine the recurrence idea
\cite{GVV_05} and the B steps in the Gottesman-Lo protocol \cite{TwoWay_03}. The
simulation results are shown in FIG.~\ref{Fig:TwoWay}.

From FIG.~\ref{Fig:TwoWay}, we can see that the recurrence scheme can increase the
key rate by around 10\% and extend the maximal tolerable loss by around 1dB. The PDC
experiment setup can tolerate up to 70dB channel loss with 3 B steps. We remark that
70dB (35dB in each channel) is comparable to the estimated loss in a satellite to
ground transmission \cite{Zeilinger_Satellite_03, Zeilinger_SpaceGround_04}.
This result suggests that satellite-ground QKD may be possible.  However, this
simulation assumes the ideal situation that an infinite number of signals are
transmitted. Moreover, we assume that $\mu$ (the brightness of the PDC source) is a
freely adjustable parameter in the PDC experiment. In a more realistic case where a
finite number of signals are transmitted and $\mu$ is a fixed parameter, the
tolerable channel loss becomes smaller, as we show next.

\subsection{Statistical fluctuations} \label{StaFlu}
In Eq.~\eqref{Post:deltabp}, we assume that $\delta_b$ and
$\delta_p$ are the same due to the symmetry between $X$ and $Z$
measurements. Alice and Bob randomly choose to measure in $X$ or $Z$
basis. Then asymptotically, $\delta_b$ is good estimate of
$\delta_p$. However, in a realistic QKD experiment, only a finite
number of signals are transmitted. Thus $\delta_p$ may slightly
differ from $\delta_b$.
We assume that Alice and Bob do not perform error testing explicitly. Instead, they
obtain the bit error rate directly from an error correction protocol (e.g., the
Cascade protocol \cite{BrassardSalvail_93}). In that case, there is no fluctuation
in the bit error rate $\delta_b=E_\lambda$. On the other hand, the phase error rate
may fluctuate to some certain value $\delta_p=\delta_b+\epsilon$. Following the
fluctuation analysis of Ref.~\cite{ShorPreskill_00}, we know that the probability to
get a $\epsilon$ bias is
\begin{equation} \label{Simulation:Confi}
P_{\epsilon} \leq \exp[{-\frac{\epsilon^2n}{4\delta_b(1-\delta_b)}}],
\end{equation}
where $n=NQ_\lambda$ the number of detection events, the product of total number of pulses $N$ and
the overall gain $Q_\lambda$.

In the 144km PDC experiment \cite{PDC144_06}, the repetition rate of pump pulse is
249MHz as given in TABLE \ref{Tab:PDC144}. As discussed in
Ref.~\cite{Zeilinger_SpaceGround_04}, the typical time of ground-satellite QKD
allowed by satellite visibility is 40 minutes. Here, we assume the experiment runs
10 minutes, which means the data size is $N=1.5\times10^{11}$. By taking this data
size, we consider the fluctuations for the entanglement PDC QKD.

\begin{figure}[hbt]
\centering \resizebox{12cm}{!}{\includegraphics{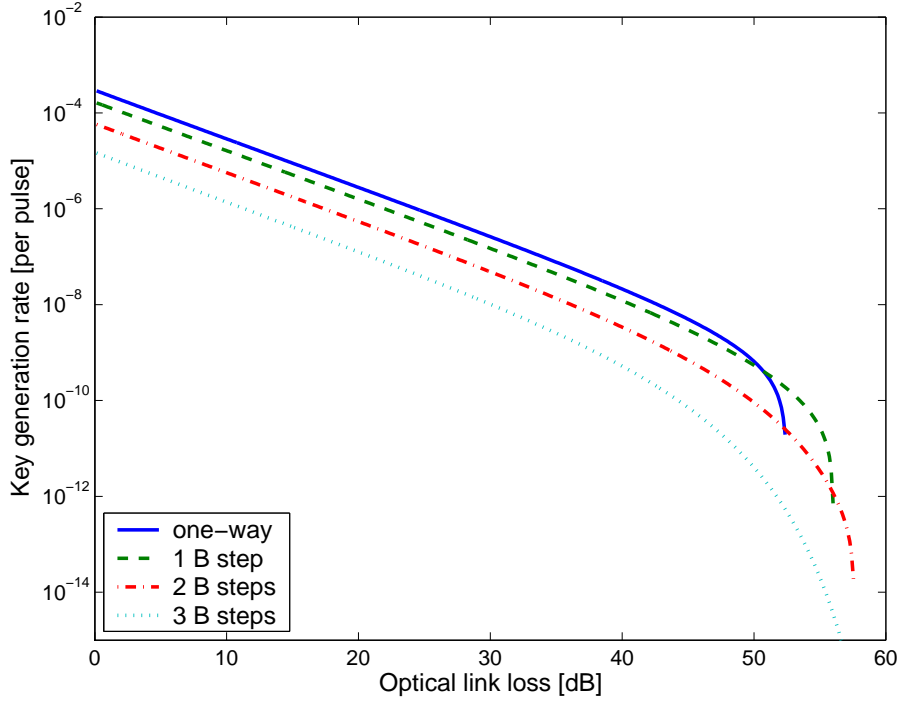}} \caption{(Color online)
Plot of the key generation rate in terms of the optical loss. We take a realistic
$\mu=2\lambda=0.053$, and consider the fluctuation with a data size of
$N=1.5\times10^{11}$ and a confident interval of $1-P_{\epsilon} \ge 1-e^{-50}$.}
\label{Fig:Fluctuation}
\end{figure}

In the realistic case, the brightness of the PDC source $\mu$ cannot be set freely.
In the 144 km PDC experiment \cite{PDC144_06}, the expected photon pair number is
$\mu=2\lambda=0.053$. After taking $\mu=0.053$ and the data size of
$N=1.5\times10^{11}$ for the fluctuation analysis, the simulation result is shown in
FIG.~\ref{Fig:Fluctuation}.

We have a couple remarks on FIG.~\ref{Fig:Fluctuation}.
\begin{enumerate}
\item
In FIG.~\ref{Fig:Fluctuation}, if we cut off from the key rate of $10^{-10}$
\footnote{Then the final key length is 15 bits. One should also consider the cost in
the authentication procedure. Thus this is a reasonable cut off point.}, the
entanglement PDC QKD with one B step can tolerate up to 53dB transmission loss.

\item
We have tried simulations with various $\mu$'s. We find that the key rate is stable
with moderate changes of $\mu$. With above fluctuation analysis, if we numerically
optimize $\mu$ for each curve, the maximal tolerable channel loss (with cut off key
rate of $10^{-10}$) is only 1dB larger than the one given by $\mu=0.053$. Thus, one
cannot significantly improve the maximal tolerable channel loss by just using a
better PDC source in the 144km PDC experiment setup \cite{PDC144_06}.


\end{enumerate}

\section{Conclusion}
We have proposed a model and post-processing for the entanglement PDC QKD. We find
that the post-processing is simple by applying Koashi-Preskill's security proof due
to the fact that the entanglement PDC QKD is a basis independent QKD. Specifically,
only directly measured data (the overall gain and the overall QBER) are needed to
perform the post-processing. By simulating a recent experiment, we compare three QKD
schemes: coherent state QKD+aysmptotic decoy, triggering PDC+asymptotic decoy, and
entanglement PDC QKD (source in the middle and at Alice's side). We find that a) the
entanglement PDC (with source in the middle) can tolerate the highest channel loss;
b) the coherent state QKD with decoy states can achieve the highest key rate in the
medium- and low-loss regions; c) asymptotically, with a realistic PDC experiment
setup, the entanglement PDC QKD can tolerate up to 70 dB channel losses by applying
post-processing schemes with two-way classical communications; d) the PDC setup can
tolerate up to 53 dB channel losses when statistical fluctuations are taken into
account.

\section{Acknowledgments}
We thank R.~Adamson,  C.~Erven, A.~M.~Steinberg and G.~Weihs for enlightening
discussions. This work has been supported by CFI, CIAR, CIPI, Connaught, CRC,
MITACS, NSERC, OIT, PREA and the University of Toronto. This research is supported
by Perimeter Institute for Theoretical Physics. Research at Perimeter Institute is
supported in part by the Government of Canada through NSERC and by the province of
Ontario through MEDT. X.~Ma gratefully acknowledges Chinese Government Award for
Outstanding Self-financed Students Abroad. C.-H. F. Fung gratefully acknowledges the
Walter C. Sumner Memorial Fellowship and the Shahid U.H. Qureshi Memorial
Scholarship.

\begin{appendix}

\section{Quantum bit error rate} \label{QBER}
Here we will study the quantum bit error rate (QBER) of the entanglement PDC QKD. Our objective is
to derive the QBER formula given in Eq.~\eqref{Model:QBER} used in the simulation. The QBER has
three main contributions:
\begin{enumerate}
\item background counts, which are random noises $e_0=1/2$;

\item intrinsic detector errors, $e_{d}$, which is the probability that a photon hit the
erroneous detector. $e_{d}$ characterizes the alignment and stability of the optical
system between Alice's and Bob's detection systems;

\item errors introduced by multi-photon-pair states: a) Alice and Bob may detect different photon
pairs; b) double clicks.  Due to the strong pulsing attack \cite{Lutkenhaus_99DoubleClick}, we
assume that Alice and Bob will assign a random bit when they get a double click. In either case,
the error rate will be $e_0=1/2$.
\end{enumerate}

Let us start with the single-photon-pair case, a Bell state given in Eq.~\eqref{Model:EPR}. The
error rate of single-photon-pair $e_1$ has two sources: background counts and intrinsic detector
errors,
\begin{equation}\label{Model:e1}
\begin{aligned}
e_1 = e_0-\frac{(e_0-e_{d})\eta_{A}\eta_{B}}{Y_1}
\end{aligned}
\end{equation}
If we neglect the case that both background and true signal cause clicks, then $e_1$ can be written
as
\begin{equation}\label{Model:e1app}
\begin{aligned}
e_1 \approx \frac{e_0(Y_{0A}Y_{0B}+Y_{0A}\eta_{B}+\eta_{A}Y_{0B})+e_{d}\eta_{A}\eta_{B}}{Y_1}.
\end{aligned}
\end{equation}
where $e_0=1/2$ is the error rate of background counts. The first term of the numerator is the
background contribution and the second term comes from the errors of true signals.

In the following, we will discuss the errors introduced by multi-photon pair states, $e_n$ with
$n\ge2$. Here we assume Alice and Bob use threshold detectors, which can only tell whether the
incoming state is vacuum or non-vacuum.  One can imagine the detection of an $n$-photon-pair state
as follows.
\begin{enumerate}

\item Alice and Bob
project the $n$-photon-pair state, Eq.~\eqref{Model:PDCn}, into $Z^{\otimes n}$ basis.

\item Afterwards, they detect each photon with certain probabilities
($\eta_A$ for Alice and $\eta_B$ for Bob).

\item If either Alice or Bob detects vacuum, then we regard it as a \emph{loss}.
If Alice and Bob both detect non-vacuum only in one polarization ($\leftrightarrow$ \emph{or}
$\updownarrow$), we regard it as a \emph{single click} event. Otherwise, we regard it as a
\emph{double click} event.
\end{enumerate}

The state of a $2$-photon-pair state, according to Eq.~\eqref{Model:PDCn}, can be written as
\begin{equation}\label{Model:PDC2pair}
\begin{aligned}
|\Phi_2\rangle&=\frac{1}{\sqrt{3}}(|2,0\rangle_a|0,2\rangle_b-|1,1\rangle_a|1,1\rangle_b+|0,2\rangle_a|2,0\rangle_b \\
&= \frac{1}{\sqrt{3}}[|\leftrightarrow\leftrightarrow\rangle_a|\updownarrow\updownarrow\rangle_b
-\frac12(|\leftrightarrow\updownarrow\rangle+|\updownarrow\leftrightarrow\rangle)_a\otimes(|\updownarrow\leftrightarrow\rangle+|\leftrightarrow\updownarrow\rangle)_b
+|\updownarrow\updownarrow\rangle_a|\leftrightarrow\leftrightarrow\rangle_b]. \\
\end{aligned}
\end{equation}
As discussed above, Alice and Bob project the state into $Z \otimes Z$ basis.
If they end up with the first or the third state in the bracket of Eq.~\eqref{Model:PDC2pair}, they
will get perfect anti-correlation, which will not contribute to errors. If they get the second state in
the bracket of Eq.~\eqref{Model:PDC2pair}, their results are totally independent, which will cause
an error with probability $e_0=1/2$. Thus the error probability introduced by a $2$-photon-pair
state is 1/6. Here we have only considered the errors introduced by multi photon states, item 3
discussed in the beginning of this Appendix.
We should also take into account the effects of background counts and intrinsic detector errors.
With these modifications, the error rate of $2$-photon-pair state is given by
\begin{equation}\label{Model:e2}
\begin{aligned}
e_2 = e_0-\frac{2(e_0-e_{d})[1-(1-\eta_A)^2][1-(1-\eta_B)^2]}{3Y_2}
\end{aligned}
\end{equation}
where $Y_2$ is given in Eq.~\eqref{Model:Yn}. Eq.~\eqref{Model:e2} can be understood as follows.
Only when Alice and Bob project the Eq.~\eqref{Model:PDC2pair} into
$|\leftrightarrow\leftrightarrow\rangle_a|\updownarrow\updownarrow\rangle_b$ or
$|\updownarrow\updownarrow\rangle_a|\leftrightarrow\leftrightarrow\rangle_b$ and no background
count occurs, they have a probability of $e_d$ to get the wrong answer. Given a coincident
detection, the conditional probability for this case is
${2[1-(1-\eta_A)^2][1-(1-\eta_B)^2]}/{3Y_2}$. All other cases, a background count, a double click
and measuring different photon pairs,
will contribute an error probability $e_0=1/2$.


Next, let us study the errors coming from the state $|n-m,m\rangle_a|m,n-m\rangle_b$. When Alice
detects at least one of $n-m$ $|\updownarrow\rangle$ photons but none of $m$
$|\leftrightarrow\rangle$ photons, and Bob detects at least one of $n-m$ $|\leftrightarrow\rangle$
photons but none of $m$ $|\updownarrow\rangle$ photons, or both Alice and Bob have bit flips of
this case, they will end up with an error probability of $e_d$. Given a coincident detection, the
conditional probability for these two cases is
\begin{displaymath}
\begin{aligned}
\frac{1}{Y_n}& \{[1-(1-\eta_A)^{n-m}](1-\eta_A)^{m}[1-(1-\eta_B)^{n-m}](1-\eta_B)^{m}\\
&+[1-(1-\eta_A)^m](1-\eta_A)^{n-m}[1-(1-\eta_B)^m](1-\eta_B)^{n-m}\}. \\
\end{aligned}
\end{displaymath}
When Alice detects at least one of $n-m$ $|\updownarrow\rangle$ polarizations but none of $m$
$|\leftrightarrow\rangle$ polarizations, and Bob detects at least one of $m$ $|\updownarrow\rangle$
polarizations but none of $n-m$ $|\leftrightarrow\rangle$ polarizations, or both Alice and Bob have
bit flips of this case, they will end up with an error probability of $1-e_d$. Given a coincident
detection, the conditional probability for these two case is
\begin{displaymath}
\begin{aligned}
\frac{1}{Y_n}&\{[1-(1-\eta_A)^m](1-\eta_A)^{n-m}[1-(1-\eta_B)^{n-m}](1-\eta_B)^{m}  \\
&+[1-(1-\eta_A)^{n-m}](1-\eta_A)^{m}[1-(1-\eta_B)^{m}](1-\eta_B)^{n-m}\}. \\
\end{aligned}
\end{displaymath}
For all other cases, the error probability is $e_0$. Thus the error probability for the state
$|n-m,m\rangle_a|m,n-m\rangle_b$ is
\begin{equation}\label{Model:errn}
\begin{aligned}
e_{nm}
=& e_0-\frac{e_0-e_{d}}{Y_n} \{(1-\eta_A)^{n-m}(1-\eta_B)^{n-m} [1-(1-\eta_A)^m][1-(1-\eta_B)^m] \\
&+(1-\eta_A)^{m}(1-\eta_B)^{m}[1-(1-\eta_A)^{n-m}][1-(1-\eta_B)^{n-m}] \\
&-(1-\eta_A)^{n-m}(1-\eta_B)^{m} [1-(1-\eta_A)^m][1-(1-\eta_B)^{n-m}] \\
&-(1-\eta_A)^{m}(1-\eta_B)^{n-m}[1-(1-\eta_A)^{n-m}][1-(1-\eta_B)^{m}]\} \\
=& e_0-\frac{e_0-e_{d}}{Y_n} [(1-\eta_A)^{n-m}-(1-\eta_A)^{m}][(1-\eta_B)^{n-m}-(1-\eta_B)^m] \\
\end{aligned}
\end{equation}

In general, for an $n$-photon-pair state described by Eq.~\eqref{Model:PDCn}, the error rate is
given by
\begin{equation}\label{Model:en}
\begin{aligned}
e_n &= \frac{1}{n+1}\sum_{m=0}^{n} e_{nm} \\
     &= \frac{1}{n+1}\sum_{m=0}^{n} e_0-\frac{e_0-e_{d}}{Y_n} [(1-\eta_A)^{n-m}-(1-\eta_A)^{m}][(1-\eta_B)^{n-m}-(1-\eta_B)^m] \\
     &= e_0-\frac{e_0-e_{d}}{(n+1)Y_n}\sum_{m=0}^{n}[(1-\eta_A)^{n-m}-(1-\eta_A)^{m}][(1-\eta_B)^{n-m}-(1-\eta_B)^m] \\
     &= e_0-\frac{2(e_0-e_{d})}{(n+1)Y_n} [\frac{1-(1-\eta_A)^{n+1}(1-\eta_B)^{n+1}}{1-(1-\eta_A)(1-\eta_B)}-\frac{(1-\eta_A)^{n+1}-(1-\eta_B)^{n+1}}{\eta_B-\eta_A}] \\
\end{aligned}
\end{equation}

The overall QBER is given by
\begin{equation}\label{Model:QBERApp}
\begin{aligned}
E_{\lambda}Q_{\lambda} =& \sum_{n=0}^{\infty} e_nY_nP(n) \\
               =& e_0Q_{\lambda}-\sum_{n=0}^{\infty} \frac{2(e_0-e_{d})\lambda^n}{(1+\lambda)^{n+2}}
               [\frac{1-(1-\eta_A)^{n+1}(1-\eta_B)^{n+1}}{1-(1-\eta_A)(1-\eta_B)}-\frac{(1-\eta_A)^{n+1}-(1-\eta_B)^{n+1}}{\eta_B-\eta_A}] \\
               =&e_0Q_{\lambda}-\frac{2(e_0-e_{d})\eta_A\eta_B\lambda(1+\lambda)}{(1+\eta_A\lambda)(1+\eta_B\lambda)(1+\eta_A\lambda+\eta_B\lambda-\eta_A\eta_B\lambda)} \\
\end{aligned}
\end{equation}
where $Q_{\lambda}$ is the gain given in Eq.~\eqref{Model:Gain}.

\section{Optimal $\mu$} \label{Optimalmu}
The optimal $\mu$ for the coherent state QKD has already been discussed
\cite{IndividualAttack_00,Practical_05}. Here we need to find out the optimal $\mu$ for the
entanglement PDC QKD. In the following calculation, we will focus on optimizing the parameter
$\lambda$ ($=\mu/2$) for the key generation rate given in Eq.~\eqref{Post:KeyEn}.

By assuming $\eta_B$ to be small and neglecting $Y_0$, we can simplify Eq.~\eqref{Model:Gain}
\begin{equation}\label{Simulation:Gainapp}
\begin{aligned}
Q_{\lambda} 
        &\approx 2\eta_B\lambda[1-\frac{1-\eta_A}{(1+\eta_A\lambda)^3}]. \\
\end{aligned}
\end{equation}

The overall QBER given in Eq.~\eqref{Model:QBER} can be simplified to
\begin{equation}\label{Simulation:QBERapp}
\begin{aligned}
E_{\lambda} 
        &\approx\frac12-\frac{(1-2e_{d})(1+\lambda)(1+\eta_A\lambda)}{2(1+3\lambda+3\eta_A\lambda^2+\eta_A^2\lambda^3)}. \\
\end{aligned}
\end{equation}

In order to maximize the key generation rate, given by Eq.~\eqref{Post:KeyEn}, the optimal
$\lambda$ satisfies
\begin{equation} \label{Simulation:KeyEndiffmu}
\frac{\partial Q_\lambda}{\partial \lambda}[1-(1+f(E_\lambda))H_2(E_\lambda)]
-Q_\lambda[1+f(E_\lambda)]\frac{\partial E_\lambda}{\partial
\lambda}\log_2\frac{1-E_\lambda}{E_\lambda}=0.
\end{equation}
Here we treat $f(E_\lambda)$ as a constant. In the following we will consider two extremes:
$\eta_A\approx1$ and $\eta_A\ll1$.

When $\eta_A\approx1$, the overall gain and QBER are given by
\begin{equation}\label{Simulation:GainappA1}
\begin{aligned}
Q_{\lambda} 
        &\approx 2\eta_B\lambda \\
E_{\lambda} 
        &\approx\frac{2e_d+\lambda}{2+2\lambda}. \\
\end{aligned}
\end{equation}
Thus, Eq.~\eqref{Simulation:KeyEndiffmu} can be simplified to
\begin{equation} \label{Simulation:KeyEndiffmuA1}
1-[1+f(E_\lambda)]H_2(E_\lambda)-\lambda[1+f(E_\lambda)]\frac{1-2e_d}{2(1+\lambda)^2}
\log_2\frac{1-E_\lambda}{E_\lambda}=0.
\end{equation}

\begin{figure}[hbt]
\centering \resizebox{12cm}{!}{\includegraphics{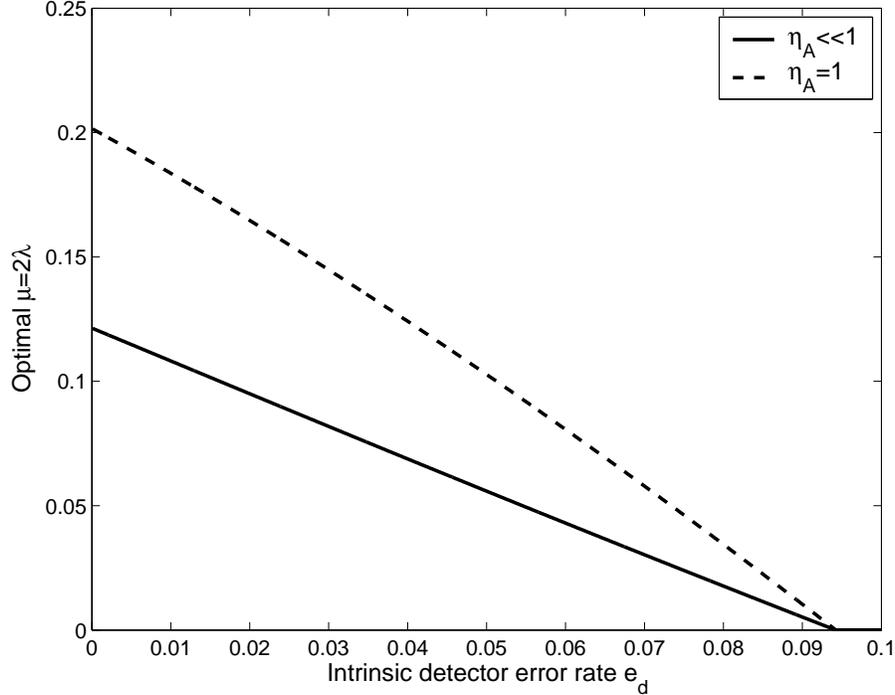}} \caption{Plot of the optimal $\mu$ in
terms of $e_d$ for the entanglement PDC QKD. $f(e_d)=1.22$.} \label{Fig:optmuen}
\end{figure}

When $\eta_A\ll1$,
\begin{equation}\label{Simulation:GainappA0}
\begin{aligned}
Q_{\lambda} 
        &\approx 2\eta_A\eta_B\lambda(1+3\lambda) \\
E_{\lambda} 
        &\approx \frac{e_d+\lambda+e_d\lambda}{1+3\lambda}.
\end{aligned}
\end{equation}
Thus, Eq.~\eqref{Simulation:KeyEndiffmu} can be simplified to
\begin{equation} \label{Simulation:KeyEndiffmuA0}
(1+6\lambda)\{1-[1+f(E_\lambda)]H_2(E_\lambda)\}-\lambda[1+f(E_\lambda)]\frac{1-2e_d}{1+3\lambda}
\log_2\frac{1-E_\lambda}{E_\lambda}=0.
\end{equation}
The solutions to Eqs.~\eqref{Simulation:KeyEndiffmuA1} and \eqref{Simulation:KeyEndiffmuA0} are
shown in FIG.~\ref{Fig:optmuen}.

From FIG.~\ref{Fig:optmuen}, we can see that the optimal
$\mu=2\lambda$ for the entanglement PDC is in the order of 1,
$\mu=2\lambda=O(1)$, which will lead the final key generation rate
to be $R=O(\eta_A\eta_B)$.

\end{appendix}

\bibliographystyle{ieeetr}

\bibliography{Bibli}


\end{document}